\documentstyle[12pt,epsfig]{article}
\def\bmq{{\bf q}}
\def\bmr{{\bf r}}
\def\bmx{{\bf x}}
\def\bmy{{\bf y}}
\def\bmz{{\bf z}}
\def\bmJ{{\bf J}}
\def\bmD{{\bf D}}
\def\bmO{{\bf O}}
\def\bmN{{\bf N}}
\def\bmmu{\mbox{\boldmath$\mu$}}
\def\bmsig{\mbox{\boldmath$\sigma$}}
\def\lsim{\mathrel{\raise3pt\hbox to 8pt{\raise -6pt\hbox{$\sim$}\hss{$<$}}}}
\def\rsim{\mathrel{\raise2pt\hbox to 9pt{\raise -7pt\hbox{$\sim$}\hss{$>$}}}}

\begin{document}

\begin{center}

{\Large {\bf Higher-Order Nuclear-Polarizability Corrections in
Atomic Hydrogen}}\\

\vspace*{0.80in}

J.L.\ Friar \\
Theoretical Division \\
Los Alamos National Laboratory \\
Los Alamos, NM  87545 \\

\vspace*{0.1in}
and

Institute for Nuclear Theory\\
University of Washington\\
Seattle, WA 98195-1550\\

\vspace*{0.20in}
and
\vspace*{0.20in}

G.\ L.\ Payne\\
Department of Physics and Astronomy\\
University of Iowa\\
Iowa City, IA 52242\\

\end{center}

\vspace*{0.50in}

\begin{abstract}
Nuclear-polarizability corrections that go beyond unretarded-dipole  
approximation are calculated analytically for hydrogenic (atomic)  
S-states. These retardation corrections are evaluated numerically for 
deuterium and contribute -0.68 kHz, for a total polarization correction 
of 18.58(7) kHz. Our results are in agreement with one previous numerical 
calculation, and the retardation corrections completely account for the 
difference between two previous  
calculations.  The uncertainty in the deuterium polarizability correction  
is substantially reduced.  At the level of 0.01 kHz for deuterium, only  
three primary nuclear observables contribute:  the electric polarizability,  
$\alpha_E$, the paramagnetic susceptibility, $\beta_M$, and the third  
Zemach moment, $\langle r^3 \rangle_{(2)}$.  Cartesian multipole  
decomposition of the virtual Compton amplitude and its concomitant  
gauge sum rules are used in the analysis.
\end{abstract}
\pagebreak

\begin{center}
{\large {\bf Introduction}}\\
\end{center}

The remarkable experiments presently being performed in Garching\cite{1} 
and Paris\cite{2} on the spectroscopy of hydrogen isotopes  
have astonishing precision.  The Rydberg currently has an uncertainty  
of 9 parts per trillion, while the isotope shift between deuterium and  
hydrogen 1S - 2S transitions has a reported uncertainty of 3 parts per  
billion, and this is expected to be lowered soon by an order of  
magnitude\cite{3}.

The isotope-shift measurements afford a unique opportunity for nuclear  
physics.  The traditional technique for determining nuclear sizes is to  
scatter relativistic electrons from nuclei, determine the charge form  
factor, and extrapolate this to small momentum transfers, thus  
determining the mean-square charge radius, $\langle r^2 \rangle$.  It is  
extremely difficult to perform the latter measurements with an absolute  
accuracy of one percent or less, and this sets limits on the accuracy of  
the charge radius.  The currently accepted value of the charge radius of  
the proton\cite{4}, $\langle r^2 \rangle^{1/2}_p = 0.862(12)$ fm,
corresponds to an uncertainty in $\langle r^2 \rangle$ of nearly 3  
percent, and the recently determined deuteron radius\cite{5}, $\langle  
r^2 \rangle_d^{1/2} = 2.128(11)$ fm, has an uncertainty in 
$\langle r^2 \rangle$ of 1 percent. For the 1S-2S d-p isotope  
shift\cite{1} the nuclear-size correction contributes approximately 
-5000 kHz (roughly the same as the QED corrections) out of a total of 
670 GHz.  The reported\cite{3} uncertainty of 2 kHz corresponds to a  
precision of better than one part per thousand in $\langle r^2 \rangle$.

In addition to static size corrections, the electron polarizes the nucleus  
and produces nuclear-polarizability corrections.  In order to use the  
isotope shift as a precise gauge of nuclear size differences, it is  
necessary to compute these polarizability corrections as accurately as  
possible, and that is the goal of this work.

There have been several calculations of these corrections for deuterium  
\cite{6,7,8,8x,9,10}.  The bulk of the effect ($\approx$19 kHz 
{\it in toto}) is caused by the Coulomb interaction distorting the  
nucleus ($\approx$17 kHz) with a smaller ($\approx$2 kHz)  
contribution from the virtual transverse photons.  In leading order  
(unretarded-dipole approximation) the electric polarizability,  
$\alpha_E$, dominates the process and accounts for 19.26(6) kHz in  
nonrelativistic approximation for the deuteron \cite{10}.  This  
numerical result summarized calculations for a group of 
``second-generation'' potentials, which fit the nucleon-nucleon scattering  
data well enough to be considered alternative phase-shift analyses of  
that data.

\begin{figure}[htb]
\epsfig{file=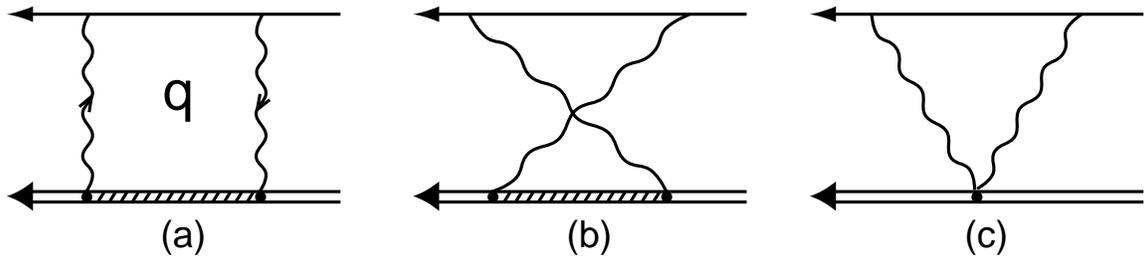,height=1.32in}
\caption{Nuclear polarization corrections with direct (a), crossed (b),
and seagull (c) contributions are illustrated. Single lines represent 
an electron, double lines a nucleus, and shaded double lines depict an excited
nucleus, with the seagull vertex maintaining gauge invariance.}
\end{figure}

There exists a single calculation \cite{7}, using first-generation nuclear  
potentials, that goes beyond the unretarded-dipole approximation and  
includes retardation, higher multipoles, the effect of the finite sizes of  
the nucleons, seagulls, and even meson-exchange currents.  Results for  
this calculation are smaller by $\rsim$ 0.5 kHz than for those using the 
unretarded-dipole approximation.  That calculation was performed by  
constructing nuclear charge and current (transition) densities and  
performing a difficult double integral over the momentum and energy  
transferred across each photon line (see Figure 1).  Our goal is to reduce  
that calculation to an analytic series in various size-dependent nuclear  
observables, and keep only those that are expected to contribute at a  
level of greater than 0.01 kHz.  The resulting expression is fairly simple 
and depends only on three primary nuclear observables:  the electric  
polarizability, $\alpha_E$, the paramagnetic susceptibility,  
$\beta_M$, and the third Zemach moment \cite{11,12}, $\langle r^3  
\rangle_{(2)}$ , of the charge distribution.  Meson-exchange currents\cite{13} 
play a small role that is easily incorporated in the calculation.  Our 
final result is in excellent agreement with the difficult but  
comprehensive calculation of Ref.\ \cite{7}.  We will produce a final  
estimate for the complete polarizability correction of 18.58(7) kHz, based 
on second-generation potentials.  It will be not be easy to improve this 
result significantly, because it will be difficult to increase the precision
of the nuclear observables on which the result depends.

\begin{center}
{\large {\bf Higher Polarizabilities}}\\
\end{center}

The integrals over momentum transfer in the loops that define the  
generalized polarizability correction are difficult, rather complicated,  
and extremely tedious.  For all these reasons, we have relegated them to  
Appendix A.  The constraints of gauge invariance are crucial to impose  
(the results are infrared divergent otherwise), but are also tedious to  
develop, although they have been known for decades\cite{14}.
Consequently, a brief presentation of the necessary relations has been  
relegated to Appendix B.  Only those parts of the calculation that we  
will treat numerically are given directly below.  To the order that we
work, gauge invariance has been properly implemented.

We first define the electric polarizability\cite{14,15}, $\alpha_E$, in terms 
of the electric-dipole operator, $\bmD$,
$$
\alpha_E = \frac{2 \alpha}{3} \sum_{N \neq 0}  
\frac{| \langle N| \bmD | 0 \rangle |^{2}}{E_{N} - E_{0}}\, , 
\eqno (1)
$$
and the logarithmic mean-excitation energy\cite{10,16}, $\bar{E}$,  by
$$
\log (\bar{E} / m_e)\, \alpha_E = \frac{2 \alpha}{3} 
\sum_{N \neq 0} \frac{| \langle N | \bmD | 0 \rangle |^{2}}{E_{N} - E_{0}} 
\log \left[ (E_N - E_0)/m_e \right]\, . \eqno (2)
$$
Similarly, we define the paramagnetic susceptibility\cite{14,15}, $\beta_M$, 
in terms of the magnetic-(dipole) moment operator, $\bmmu$, (see Eq.\ (B8))
$$
\beta_M = \frac{2 \alpha}{3} \sum_{N \neq 0}  
\frac{| \langle N| \bmmu | 0 \rangle |^{2}}{E_{N} - E_{0}}\, , 
\eqno (3)
$$
together with its mean-excitation energy, $\bar{E}$,
$$
\log (\bar{E} / m_e)\, \beta_M = \frac{2 \alpha}{3} 
\sum_{N \neq 0} \frac{| \langle N | \bmmu | 0 \rangle |^{2}}{E_{N} - E_{0}} 
\log \left[ (E_N - E_0)/m_e \right]\, . \eqno (4)
$$
A close relative of $\alpha_{E}$ is
$$
\Delta \alpha_E = \frac{2 \alpha}{3} \sum_{N \neq 0}  
\frac{| \langle N| \bmD | 0 \rangle |^{2}}{(E_{N} - E_{0})^3} \, , 
\eqno (5)
$$
together with its logarithmic mean-excitation energy analogously  
defined.  Although we have used (and will write below) $\bar{E}$ for  
each of the {\it different} mean-excitation energies for  
notational simplicity, they are distinct (although similar) numbers.   
They will always be grouped with the operators that define them.  In  
the formulae above, $\alpha$ is the fine-structure constant, $m_e$ is  
the electron mass, $|N \rangle$ is the N\underline{th} nuclear state
(the N\underline{th} eigenstate of $H_0$) with energy, $E_N$, and $N = 0$ 
labels the ground state.

The inelastic charge density (squared) can be rewritten in cases where  
there are no energy factors\cite{17}
$$
\sum_{N \neq 0} \langle 0| \rho^{\dagger} (\bmx) | N \rangle 
\langle N | \rho (\bmy) | 0 \rangle = \langle 0 | \rho^{\dagger} (\bmx ) 
\rho (\bmy) | 0 \rangle - \rho_0 (x) \rho_0 (y) \, ,  \eqno (6)
$$
where
$$
\rho_0 (x) = \langle 0 | \rho (\bmx ) | 0 \rangle \, , \eqno (7)
$$
$\rho (\bmx) $ is the nuclear charge (-density) operator 
and $\int d^3 x\, \rho_0 (x) = Z$, the nuclear charge.

With these definitions, we can rewrite Eq.\ (A30) in the form
\begin{eqnarray*}
\lefteqn{\Delta E_n = -4 \alpha\, m_e | \phi_n (0)|^2 
\left[ \frac{5 \alpha_E}{4} 
[\log (2\bar{E} / m_e) + \frac{19}{30}] + \frac{15}{16} m_e^2\, \Delta \alpha_E 
[\log (2\bar{E} / m_e) - \frac{283}{300}] \right.} \\
&-&\frac{\beta_M}{4}[\log (2\bar{E} / m_e) - \frac{1}{6}]
+\frac{\pi \alpha}{12} \int d^3 x \int d^3 y\, |\bmx - \bmy|^3\, 
\left[ \langle 0 | \rho^{\dagger} (\bmx ) 
\rho (\bmy) | 0 \rangle - \rho_0 (x) \rho_0 (y) \right] \\
&+& \left. \Delta B \rule{0in}{3ex} \right] \, , \hspace*{4.9in}  (8)
\end{eqnarray*}
where
$$
\Delta B = \frac{\alpha}{5}\left( \gamma + 
\log \left( \bar{E}\bar{z} \right)- \frac{39}{20}\right) 
\int d^3 x \int d^3 y\; \langle 0 | \Delta B (\bmx , \bmy ) | 0 \rangle
\, ,  \eqno (9)
$$
and
$$
\Delta B (\bmx , \bmy ) = B^{ii}_{\rm in} (\bmx , \bmy ) 
[\bmx \cdot \bmy  - y^2 ]
+ B^{ij}_{\rm in} (\bmx , \bmy ) [ -2 y^i y^j + x^i y^j + y^i x^j ] + \cdots
\, . \eqno (10)
$$
We have dropped a large variety of small terms in $\Delta B$ (indicated 
by dots and with relative coefficients $\sim$1/10) that arise  
from the seagull and current terms (converted using gauge sum rules;  
see Appendix A).  Gauge sum rules and approximating $E_N - E_0$  
by $\bar{E}$ and $z \equiv | \bmx -\bmy |$ by $\bar{z}$ (i.e., constants) 
were used to obtain Eqs.\ (9) and (10). The small correction $\Delta B$ arises
from the Coulomb interaction; we will also use the alternative expression in
Eq.\ (A31). Note that the term $\sim | \bmx - \bmy|^3$ contributes  
to {\it all} electric multipoles, unlike the others, which are either
electric- or magnetic-dipole in nature.  Only ground-state properties  
are needed to construct the former term.

In order to proceed further, we need to specify our nuclear model.  The  
charge operator $\rho (\bmx)$ is given by
$$
\rho (\bmx ) = \sum_{i = 1}^A \hat{e}_i (\bmx - \bmx_i )\, , \eqno (11)
$$
where
$$
\hat{e}_i = \hat{e}_p (i) \; e_p (\bmx - \bmx_i ) + \hat{e}_n (i)\; 
e_n (\bmx - \bmx_i ) \eqno (12)
$$
counts protons (with $\hat{e}_p$) and neutrons (with $\hat{e}_n$) and  
multiplies each species by its intrinsic charge distribution.  The form  
(11) is nothing more than the usual folding of $\hat{e}_i (x)$ with  
$\delta^3 (\bmx - \bmx_i)$.  Forming $\int d^3 x\, \bmx\, \rho (\bmx) = 
\sum_i \hat{e}_p (i) \bmx_i$ demonstrates that finite size  
does not modify the electric-dipole operator.

The current operator needed to construct $\bmmu$ consists of three distinct
parts: the spin-magnetization current, the orbital current, and 
meson-exchange currents (MEC). We ignore nucleon finite size, which does not 
contribute to this order, and find\cite{18}
$$
\bmmu = \sum_{i = 1}^A \left( \frac{\hat{\mu} (i) \, \bmsig (i) + 
\hat{e}_p (i) \, {\bf L} (i)}{2 M} \right) + \bmmu_{MEC} \, , \eqno (13)
$$
where the spin-magnetization current is determined by
$$
\hat{\mu} (i) = \mu_p \, \hat{e}_p (i) + \mu_n \, \hat{e}_n (i) \, . \eqno (14)
$$
Note that the isoscalar and isovector nucleon magnetic moments are very
different in size: $\mu_s \equiv \mu_p + \mu_n = 0.8798\ldots$ and 
$\mu_v \equiv \mu_p - \mu_n = 4.7059\ldots$.  The large value of the isovector 
nucleon magnetic moment will play a determinative role.  We eschew writing out  
our model for the two-body pion-exchange currents(i.e., MEC), which is 
discussed in Ref.\ \cite{18}.  This model has had its pion-nucleon form factor 
adjusted to reproduce the experimental thermal $n - p$ radiative capture rate
\cite{19}.  As the contribution of $\beta_M$ is relatively small and the  
MEC a small part of this, the overall MEC contribution is nearly  
negligible, but has been included for completeness.

Our final ingredient is the Compton seagull operator.  This operator is  
comprised of several components\cite{20}:  impulse approximation, plus  
meson-exchange currents\cite{21}, plus \ldots.  We expect the meson-exchange  
currents to be possibly comparable to the impulse approximation,  
based on sum-rule studies\cite{22}.  We will work only with the  
impulse-approximation component, which has the form\cite{14}
$$
B^{ij} (\bmx , \bmy ) = \sum_{i = 1}^A \hat{e}_i (\bmx - \bmx_i )\;
\hat{e}_i (\bmy - \bmx_i )\, . \eqno (15)
$$
The pion-exchange component of the deuteron's diamagnetic susceptibility (see  
Eq.\ (A29) and Refs.\ \cite{20,21}) has been shown to be tiny.

\begin{center}
{\large {\bf Numerical Calculations}}
\end{center}

Podolsky's method\cite{23} is very convenient for calculating $\alpha_E$  
and $\beta_M$.  Any generalized polarizability of the type displayed in  
Eqs.\ (1) and (3) can be calculated as follows.  Equation (1) is fully  
equivalent to
$$
\alpha_{E} = 2 \alpha \; \langle 0 | D_{z} | \Delta \Psi_z \rangle 
\, , \eqno (16)
$$
where
$$
(H_0 - E_0) \; | \Delta \Psi_z \rangle \; = \; D_z | 0 \rangle \eqno (17)
$$
is solved subject to finite boundary conditions.  One must be careful to 
exclude the ground state from the sum over N in Eq.\ (3) for $\beta_M$.
This necessitates a projection orthogonal to the ground state on the  
right-hand side of Eq.\ (17) (with $D_z$ replaced by $\mu_z$ in both Eqs.\ (16) 
and (17)).

For the deuteron, one impulse-approximation calculation of $\beta_M$ 
exists\cite{20} with  
a value of 0.065 fm$^3$.  This is dominated by $^1\!S_0$ intermediate 
states.  Indeed, an upper limit for all triplet intermediate states is 
$\beta^t_M \leq 0.0003\, {\rm fm}^3$, obtained\cite{20} by replacing the  
energy denominator in Eq.\ (3) by its smallest possible value ($E_d$, the 
deuteron binding energy), and then using closure and completing the algebra.  
Moreover, the $^1\!D_2$-state contribution is tiny, and the $^1\!S_0$ 
intermediate states dominate completely.

The logarithmic mean-excitation energies are calculated using a  
trick\cite{24}.  We define a quantity closely related to $\alpha_E$
$$
\alpha_E (\xi) = \frac{2 \alpha}{3} \; \sum_{N \neq 0} \; \frac{| 
\langle N | \bmD | 0 \rangle |^2}{\xi f + E_N - E_0}\, , \eqno (18)
$$
where $f$ is an energy-scaling factor $\sim$ 3-5$E_d$ inserted for  
convenience.  An integral over $\xi$ produces a logarithm, and one  
finds a convenient numerical algorithm for $\bar{E}$ in
$$
\alpha_E (0) \, \log \; (2 \bar{E}/m_e) = \int^{1}_{0} \; 
\frac{d \xi}{\xi} [ \alpha_E (\xi) - \alpha_E (0) + \alpha_E 
( 1 / \xi )] - \alpha_E (0)\, \log (m_e/2 f) \, , \eqno (19)
$$
where $\bar{E}$ is independent of $f$, and Eq.\ (19) is fully  
equivalent to Eq.\ (2). 

The electric polarizability was calculated and thoroughly discussed in  
Ref.\ \cite{10}.  One found there
\begin{eqnarray*}
\hspace*{2in} \alpha_E &=& 0.6328(17)\, {\rm fm}^3\\
\log (2 \bar{E} / m_e )&=& 2.9620(5) \hspace*{2.4in} (20)\\
\nu_{\rm pol}^{\alpha_E} &=& 19.26(6) \, {\rm kHz} = 
(16.98) + (2.28)\;{\rm kHz}\, ,
\end{eqnarray*}
where the latter is broken down into Coulomb, transverse, and total  
contributions.  This calculation did not incorporate relativistic  
corrections to the deuteron.

\begin{table}[htb]
\centering

\caption{Impulse-approximation deuteron magnetic susceptibilities, 
$\beta_{\rm M}$, in units of 
fm$^3$, logarithmic mean-excitation-energy ratios, $\log(2 \bar{E}/m_e)$, 
and corresponding deuteron 1S-2S polarization-energy shifts, $\nu_{\rm pol}$, 
in kHz. The RSC potential labelled [-23] has had its $^1\!S_0$ part modified 
to produce the correct n-p scattering length.}

\hspace{0.25in}

\begin{tabular}{|l|ccc|} \hline

\multicolumn{1}{|c|}{\rule{0in}{3ex} Potential Model}&
\multicolumn{1}{c}{$\beta_{\rm M} ({\rm fm}^{3})$}&
\multicolumn{1}{c}{$\log(2 \bar{E}/m_e)$}&
\multicolumn{1}{c|}{$\nu_{\rm pol} ({\rm kHz})$}\\[0.5ex] \hline \hline
\multicolumn{4}{|c|}{Second-Generation Potentials}\\ \hline

Argonne V$_{18}$ \rule{0in}{2.5ex} & 0.0678  & 2.4724  &  -0.265 \\
Nijmegen (loc-rel)    & 0.0677  & 2.4726  &  -0.264 \\
Nijmegen (loc-nr)     & 0.0677  & 2.4732  &  -0.264 \\
Nijmegen (nl-rel)     & 0.0677  & 2.4726  &  -0.264 \\
Nijmegen (nl-nr)      & 0.0676  & 2.4732  &  -0.264 \\
Nijmegen (full-rel)   & 0.0675  & 2.4728  &  -0.264 \\ 
Reid Soft Core (93)   & 0.0674  & 2.4744  &  -0.264 \\ \hline

\multicolumn{4}{|c|}{First-Generation Potentials} \\ \hline

Bonn (CS) \rule{0in}{2.5ex}& 0.0682 & 2.4738 & -0.267  \\
Argonne V$_{14}$         & 0.0674 & 2.4733 & -0.263  \\
Reid Soft Core (68)[-23] & 0.0669 & 2.4748 & -0.261  \\
Nijmegen (78)            & 0.0663 & 2.4947 & -0.261  \\
Super Soft Core (C)      & 0.0659 & 2.4982 & -0.260  \\ 
de Tourreil-Rouben-Sprung& 0.0656 & 2.4969 & -0.259  \\
Paris                    & 0.0653 & 2.5008 & -0.258  \\
Reid Soft Core (68)      & 0.0647 & 2.5031 & -0.256  \\ \hline 

\end{tabular}
\end{table}

\begin{table}[hbt]
\centering

\caption{Full deuteron magnetic susceptibilities, $\beta_{\rm M}$, in units of 
fm$^{3}$, logarithmic mean-excitation-energy ratios, $\log(2 \bar{E}/m_e)$, 
and corresponding deuteron 1S-2S polarization-energy shifts, $\nu_{\rm pol}$, 
in kHz. The RSC potential labelled [-23] has had its $^1\!S_0$ part modified 
to produce the correct n-p scattering length.}

\hspace{0.25in}

\begin{tabular}{|l|ccc|} \hline

\multicolumn{1}{|c|}{\rule{0in}{3ex} Potential Model}&
\multicolumn{1}{c}{$\beta_{\rm M} ({\rm fm}^{3})$}&
\multicolumn{1}{c}{$\log(2 \bar{E}/m_e)$}&
\multicolumn{1}{c|}{$\nu_{\rm pol} ({\rm kHz})$}\\[0.5ex] \hline \hline
\multicolumn{4}{|c|}{Second-Generation Potentials}\\ \hline

Nijmegen (full-rel) \rule{0in}{2.5ex}   & 0.0780  & 2.5003  &  -0.308 \\ 
Nijmegen (nl-rel)     & 0.0778  & 2.4981  &  -0.307 \\
Nijmegen (nl-nr)      & 0.0777  & 2.4987  &  -0.307 \\
Nijmegen (loc-rel)    & 0.0775  & 2.4972  &  -0.306 \\
Nijmegen (loc-nr)     & 0.0774  & 2.4978  &  -0.306 \\
Reid Soft Core (93)   & 0.0775  & 2.5005  &  -0.306 \\
Argonne V$_{18}$      & 0.0774  & 2.4963  &  -0.305 \\ \hline

\multicolumn{4}{|c|}{First-Generation Potentials} \\ \hline

Argonne V$_{14}$ \rule{0in}{2.5ex} & 0.0774 & 2.4996 &  -0.306  \\
Reid Soft Core (68)[-23] & 0.0769 & 2.5002 &  -0.304  \\
Bonn (CS)                & 0.0766 & 2.4935 &  -0.302  \\
Nijmegen (78)            & 0.0751 & 2.5172 &  -0.299  \\
de Tourreil-Rouben-Sprung& 0.0748 & 2.5221 &  -0.298  \\
Super Soft Core (C)      & 0.0747 & 2.5218 &  -0.298  \\ 
Paris                    & 0.0743 & 2.5253 &  -0.297  \\
Reid Soft Core (68)      & 0.0742 & 2.5295 &  -0.297  \\ \hline 
                
\end{tabular}
\end{table}

The analogous calculations for $\beta_M$ are listed in Tables I and II for 
a variety of first- and second-generation 
potentials\cite{25,26,27,28,29,30,31,32,33,34}.  The results of Table I are  
for the impulse-approximation magnetic moment (no MEC), while those of Table  
II incorporate MEC, as well.  The latter increases the former by  
approximately 15\%, which is quite typical for isovector transitions.   
We average the various second-generation results and estimate
\begin{eqnarray*}
\hspace{2in} \beta_M &=& 0.0777(3)\, {\rm fm}^{3} \\
\log (2 \bar{E} / m_e ) &=& 2.498(2) \hspace*{2.45in} (21) \\
\nu_{\rm pol}^{\beta} &=& -0.307(2)(6) \, {\rm kHz} \, .
\end{eqnarray*}
These uncertainties do not include uncertainties in the MEC, which are  
possibly 1-2\% of the total result (this is subjective); the latter is  
reflected in the second error (6) in the last Eq.\ (21).  This polarization  
contribution is nonnegligible only because $\mu_v = 4.7$; a more  
``normal'' size $\sim$1 would reduce the contribution by a factor of  
$\sim$25. Our $\nu_{\rm pol}$ is in reasonable agreement with the
zero-range result of Ref.\ \cite{8x}.

The very small corrections, $\Delta \alpha_E$, can be accurately estimated 
in zero-range approximation (which we used as a tool in Ref.\ \cite{10}).  
We first calculate $\alpha^0_E (\xi)$ using Eq. (18) and find
$$
\alpha^0_E (\xi) = \frac{\alpha\, \mu A^2_s}{12 \kappa^3} 
\frac{(\kappa^2 + \bar{\kappa}^2 + 4 \kappa \bar{\kappa})}
{(\kappa + \bar{\kappa})^4}\, , \eqno (22)
$$
where the asymptotic (reduced) s-state wave function of the deuteron has the 
form: $A_s\, e^{- \kappa r}$.  Moreover, 
$\bar{\kappa}^2 \equiv \kappa^2 (1 + \xi f)$, 
and $\mu$ is the n-p reduced mass.  Performing two derivatives leads to
$$
m^2_e\, \Delta \alpha^0_E = \frac{7 \alpha^0_E\, m^2_e}{24 E^2_d}
= 0.0098\; {\rm fm}^3 \, , \eqno (23)
$$
and is accurate to better than 1/2\%. The following integral (similar to 
Eq.\ (19)) produces the logarithmic mean-excitation energy
\begin{eqnarray*}
\hspace*{0.25in} \log \; (2 \bar{E}/m_e) &=& \frac{1}{f^2 \Delta \alpha^0_E} 
\left[ \int^{1}_{0} \; \frac{d \xi}{\xi^3} [ \alpha^0_E (\xi) - \alpha^0_E (0) 
-\xi \alpha_E^{0\, \prime} (0) - \frac{\xi^2}{2} 
\alpha_E^{0\, \prime \prime} (0) ] \right. \\
&+& \left. \int^{1}_{0} \; d \xi [ \xi \alpha^0_E ( 1 / \xi ) - 
\xi \alpha^0_E (0) - \alpha_E^{0\, \prime} (0)]\rule{0in}{3.5ex} \right] 
\, - \log (m_e/2 f)\\
&=& 2.648\, . \hspace{3.75in} (24)
\end{eqnarray*}
These results produce a total correction from $\Delta \alpha_E$: 
$$
\nu^{\Delta \alpha_E}_{\rm pol} = 0.106 \, {\rm kHz}\; = (0.060) + (0.046) \, 
{\rm kHz}\, , \eqno(25)
$$
which has been broken down into Coulomb and transverse parts, respectively.

The remaining large quantity in Eq.\ (8) is the retardation correction  
proportional to $\rho_{\rm in} (\bmx , \bmy)$.  Using Eq.\ (11), we  
expand that result and find
$$
\rho^{\dagger}(\bmx ) \rho (\bmy) = \sum_{i \neq j} \hat{e}_i (\bmx - \bmx_i )
\hat{e}_j (\bmy - \bmx_j ) + \sum_{i} \hat{e}_p (i) e_p (\bmx - \bmx_i )
e_p (\bmy - \bmx_i ) + \hat{e}_n (i) e_n (\bmx - \bmx_i ) e_n (\bmy - \bmx_i )
\, . \eqno (26)
$$
Note that if the neutron's charge distribution is set to zero, the first  
term vanishes for the deuteron (one nucleon must be a neutron), and  
only the second term survives.  Shifting the $\bmx$ and $\bmy$ integrals each 
by $\bmx_i$ in that case produces $\langle r^3 \rangle^p_{(2)}$, while the
$\rho_0 (x) \rho_0 (y)$ term generates $\langle r^3 \rangle^d_{(2)}$, and
a retardation correction
$$
\frac{\pi \alpha}{12} \left[ - \langle r^3 \rangle^d_{(2)} + 
\langle r^3 \rangle^p_{(2)} \right] \equiv \frac{\pi \alpha}{12} \Delta 
\langle r^3 \rangle_{(2)} \, , \eqno (27)
$$
where
$$
\langle r^n \rangle_{(2)} = \int d^3 r\, r^n\, \rho_{(2)} (r)\, , \eqno (28)
$$
and
$$
\rho_{(2)} (r) = \int d^3 z\, \rho (|\bmz - \bmr |)\, \rho (z) \equiv 
\rho \otimes \rho \eqno (29)
$$
is the convoluted (Zemach\cite{11,12}) density.  In Eq.\ (27), the first of the
third moments is calculated with respect to the total deuteron (including 
the finite size of the proton) convoluted charge density and the second with
the proton's convoluted density, $\rho_p$. In what follows below we will 
specialize to the deuteron, and denote by $\rho_d$ the deuteron's 
ground-state charge density (called $\rho_0$ before).

For completeness, we include the neutron contributions as well.  There  
will be a two-body correlation term (first term in Eq.\ (26)) involving  
$\rho_n \otimes \rho_p \otimes \bar{\rho}_d$, where $\bar{\rho}_d$ is  
slightly modified to account for the vector $\bmr$ specifying a correlation,  
while $\bmr/2$ determines the charge density, $\rho_d$.  In addition to  
this term, the folded proton density in Eq.\ (27) is replaced by $(\rho_p  
\otimes \rho_p + \rho_n \otimes \rho_n)$ and the deuteron charge  
density is defined by $(\rho_p + \rho_n) \otimes \rho^0_d$, where  
$\rho^0_d$ is determined by the deuteron wave function alone.

We use a simplified model of the neutron and proton form factors.  The  
proton form factor is taken to have a dipole form with the correct  
radius\cite{4} (0.862 fm).  The neutron form factor is that dipole times  
$q^2$, adjusted overall to match the experimental charge radius of 
-0.338~fm\cite{35}.

Calculations of the various moments are performed by first calculating the
deuteron density and then generating a spline fit of it. Then folding is 
performed and that convoluted density is similarly fit. Moments are calculated
ultimately with respect to the final fitted density.

\begin{table}[htb]
\centering

\caption{Zemach-moment contribution $\Delta \langle r^3 \rangle_{(2)}$, to
the Coulomb-induced retardation correction in units of fm$^{3}$, and 
corresponding deuteron 1S-2S polarization-energy shifts, $\nu_{\rm pol}$, 
in kHz.}

\hspace{0.25in}

\begin{tabular}{|l|cc|} \hline

\multicolumn{1}{|c|}{\rule{0in}{3ex} Potential Model}&
\multicolumn{1}{c}{$\Delta \langle r^3 \rangle_{(2)} \, ({\rm fm}^{3})$}&
\multicolumn{1}{c|}{$\nu_{\rm pol} ({\rm kHz})$}\\[0.5ex] \hline \hline
\multicolumn{3}{|c|}{Second-Generation Potentials}\\ \hline

Argonne V$_{18}$  \rule{0in}{2.5ex} &  -37.45  &  -0.485 \\
Reid Soft Core (93)   &  -37.44  &  -0.484 \\
Nijmegen (loc-nr)     &  -37.38  &  -0.484 \\
Nijmegen (loc-rel)    &  -37.34  &  -0.483 \\
Nijmegen (nl-rel)     &  -37.36  &  -0.483 \\
Nijmegen (nl-nr)      &  -37.31  &  -0.483 \\
Nijmegen (full-rel)   &  -37.20  &  -0.481 \\ \hline 

\multicolumn{3}{|c|}{First-Generation Potentials} \\ \hline

Super Soft Core (C) \rule{0in}{2.5ex}       &  -38.45 & -0.498 \\ 
Nijmegen (78)            &  -38.27 & -0.495 \\
Argonne V$_{14}$         &  -38.01 & -0.492 \\
de Tourreil-Rouben-Sprung&  -37.70 & -0.488 \\
Paris                    &  -37.55 & -0.486 \\
Bonn (CS)                &  -37.44 & -0.484 \\
Reid Soft Core (68)      &  -36.87 & -0.477 \\ \hline 
                
\end{tabular}
\end{table}

The results for various models (including the effect of neutrons) are  
listed in Table 3.  The neutrons lower the result by approximately 1\%.   
The second-generation results for this retardation (and higher-multipole) 
correction can be summarized by
$$
\Delta \langle r^3 \rangle_{(2)} = -37.32(12)\; {\rm fm}^3 \, , \eqno (30)
$$
and
$$
\nu_{\rm pol}^{\rm ret} = -0.483(2)\, {\rm kHz}\, . \eqno (31)
$$
This process is the only one that we will consider where higher multipoles,  
retarded $E1$, and nucleon finite size contribute.

The final task will be to estimate the size of $\Delta B$.  The  
``natural'' size of terms with numerical coefficients $\sim$1 is 
$\Delta B \sim \frac{\alpha \langle r^2 \rangle_d}{M}$, while the coefficient  
$4 m_e \alpha^2 | \phi_n (0)|^2$ has a value 0.05 kHz/fm$^3$.  Since  
$\frac{\langle r^{2} \rangle_d}{M} \sim 0.8\, {\rm fm}^3$, the natural size  
is 0.04 kHz.  We use $\bar{E} \rsim$ 10 MeV, $\bar{z}  
\rsim$ 4 fm to estimate logarithms and Eqs.\ (15) and (B12) for the  
seagull operator.  The seagull contribution has a size $\sim \frac{6}{48}$
(0.04 kHz) $\sim$ 0.005 kHz.  The higher-order current terms are of  
similar size, but largely cancel, leaving a very tiny residue.  The  
higher-order charge terms are dominated by quadrupole excitations and
have a nominal size $\sim \frac{4}{4}$ (0.04 kHz) $\sim$ 0.04 kHz,  
which is almost as large as the $\Delta \alpha_E$ Coulomb 
term.  It can be shown, however, that this size is an  
artifact, caused by neglecting some recoil terms. We must perform the estimate 
more carefully.

If one uses instead the representation in Eq.\ (A31) for the $z^4$  
charge terms, and evaluates the double 
commutators, one finds that the kinetic-energy part of $H_0$ vanishes in  
the point-nucleon limit, and otherwise has a rough size $\sim$ 0.006 kHz.  One  
can also evaluate the potential part of the commutator and find $\sim$  
0.003 kHz. These corrections are not only small, but the comparable sizes
of kinetic and potential terms are in accordance with expectations.

\begin{center}
{\large {\bf Results and Conclusions}}
\end{center}

\begin{table}
\centering
\caption{Contributions in kHz to the deuteron-polarizability frequency shift for
the 1S-2S transition together with their respective origins, separated into
Coulomb and transverse, and electric dipole, magnetic dipole, and
higher-multipole and retardation terms.}

\hspace{0.25in}

\begin{tabular}{|l||rrrrr|r|}
\hline
{Origin} &$\alpha_E$ & $\Delta \alpha_E$ & $\beta_M$ & 
$\Delta \langle r^3 \rangle_{(2)}$ & $\Delta B$ & Total\\ \hline \hline
{Coulomb} & 16.98 & 0.06 & -\hspace{0.1in} & -0.48\hspace{0.2in} 
& $<$ 0.01 & 16.56\\ 
{Transverse}& 2.28 & 0.05 & -0.31 & -\hspace{0.3in} & $<$ 0.01 & 2.02\\ \hline
{Total} & 19.26 & 0.11 & -0.31 & -0.48\hspace{0.2in} & $\lsim$ 0.01 & 18.58\\ \hline

\end{tabular}
\end{table}

Our final results are tabulated in Table 4, with breakdowns according  
to their origin.  The total for the 1S-2S transition in deuterium is
$$
\nu_{\rm pol} = 18.58(7) \, {\rm kHz} \, . \eqno (32)
$$
This is 0.68 kHz less than the leading-order $(\alpha_E)$ result, and is  
consistent with the differences between the results of Refs.\ \cite{7} and 
\cite{8}. The complete numerical results of Ref.\ \cite{7} for four 
first-generation potentials (Paris\cite{30}, $AV_{14}$\cite{32},
Nijmegen\cite{33},and Bonn CS\cite{29}) are in agreement with our results 
within 0.02 kHz for Coulomb  
and transverse parts, which must be regarded as virtually perfect  
agreement.  We note that substantially improving the uncertainty in  
Eq.\ (32) will be difficult, because it would entail substantial  
improvements\cite{10,36} in the nuclear parameter, $A_s$.

A wide variety of physical mechanisms contribute to the final result.   
Unretarded $E1$ photons (both longitudinal and transverse) generate  
the electric polarizability.  The paramagnetic susceptibility generates a  
nonnegligible term only because the nucleon isovector magnetic  
moment is nearly 5.  This term is thus $\sim$25 times larger than if it  
were of ``normal'' size.  A retardation correction contributing to all  
electric multipoles is moderately important, and is the only one of our  
terms to which the nucleon form factors contribute.  Finally, higher-order 
terms, including the seagull contribution, are estimated and shown to be tiny.

\noindent {\bf Acknowledgements}

The work of J.\ L.\ Friar was performed under the auspices of the United  
States Department of Energy, while that of G.\ L.\ Payne was supported  
in part by the United States Department of Energy. We would like to thank
Winfried Leidemann, Don Sprung, and Joan Martorell for their generous help
in providing information about their calculations.

\begin{center}
\Large{{\bf Appendix A}}\\
\end{center}

We wish to evaluate the contributions of Figures (1a)-(1c) to the
energy of the n\underline{th} hydrogenic S-state. Because this calculation 
has been set up before\cite{7}, we sketch that part of the derivation.

The nuclear energy and momentum scales are much greater than those  
of an atom.  Consequently, such large momenta flow through the photon and 
electron propagators in Fig.\ (1) that only the shortest-range part of the  
electron wave functions, $|\phi_n (0)|^2$, contributes to leading  
order in $Z \alpha$, the product of the nuclear charge, $Z$, and the  
fine-structure constant, $\alpha$.  Consequently, the momentum in  
both photon propagators is taken to be $q$ (differences in these  
momenta lead to higher-order terms in $Z \alpha$).  It is important to  
enforce the constraints of gauge invariance\cite{14} on the nuclear part of the  
interaction (the virtual Compton amplitude), and this is most conveniently 
handled using Coulomb gauge, which isolates the nuclear charge density 
from the transverse parts of the currents (Figs.\ (1a) and (1b)) and seagull  
(Fig.\ (1c)).  The calculation requires a relativistic treatment of the  
electron (since $q \gg m_e$, the electron mass), but a  
nonrelativistic (i.e., leading-order) treatment of the nucleus suffices.   
We expand the nuclear current, $J^{\mu}$, in powers of $1/M$, the nucleon  
mass, and keep no powers higher than linear.  One immediate  
consequence of the latter is the lack of (nuclear) momentum  
dependence in the nuclear charge density, $\rho (\bmx)$ (unlike the  
current density, $\bmJ (\bmx)$), and the nuclear seagull density\cite{14},
$B^{mn} (\bmx , \bmy)$, which also lacks charge components (i.e.,
$B^{\mu \nu} = 0$ for $\mu = 0$ or $\nu = 0$).  Our conventions  
follow Ref.\ \cite{37}, and correspond to natural units $(\hbar = c = 1)$.
We will incorporate meson-exchange  
currents where required by gauge invariance, although their numerical  
contribution is ultimately small (see main body of paper).  We ignore  
nuclear recoil corrections ($\sim 1/M_t$, the total nucleus mass) in  
the nuclear operators, but maintain them in reduced-mass factors in the  
atomic basis states.

The energy shift of the n\underline{th} hydrogenic S-state due to  
nuclear polarization is most conveniently calculated by performing the  
contour integral over the time component of $q^{\mu} (q_0)$ in the loops 
of Fig.\ (1), which leads to\cite{7} \\
\begin{eqnarray*}
\hspace*{.25in} \Delta E_{pol} 
&=& \frac{-2 \alpha^2 m_e}{\pi}  
| \phi_n (0) |^2 \int d^3 q \left[ \sum_{N \neq 0} 
\left[ \frac{(2E + \omega_N ) | \langle N| \rho (\bmq) |0 \rangle |^2}{E q^4
[(E + \omega_N)^2 - m^2_e]} \right. \right. \\ 
&+& \left. \left( \left[ \frac{q^2}{4 m^2_e} \right] 
\frac{2 E + \omega_N}{E q^4 [ ( E + \omega_N )^2 - m^2_e]}  
-\frac{(2 q + \omega_N)}{4 m^2_e q^3 (q + \omega_N)^2}
\right) | \langle N | \bmJ_{\bot} (\bmq) | 0 \rangle |^2 \right] \\  
&+&  \left. \frac{B^{ii \bot}_{\rm in} (\bmq)}{8 q^2 m^2_e}  
\left( \frac{1}{q} - \frac{1}{E} \right) \right], \hspace*{3.10in}  
(A1) 
\end{eqnarray*}      
where $q^2 \equiv \bmq^2, E = \sqrt{q^2 + m^2_e}$, $\omega_N =  
E_N - E_0$, the energy of excitation (relative to the ground state) of  
the N\underline{th} nuclear state (which by assumption cannot be the  
ground state).  In addition
\begin{eqnarray*}
\hspace*{1.80in} \rho(\bmq) &=& \int d^3 x \, e^{i \bmq \cdot \bmx}  
\rho (\bmx)\, , \hspace*{2.0in} (A2) \\
\hspace*{1.80in} \bmJ (\bmq) &=& \int d^3 x \, e^{i \bmq \cdot \bmx}  
\bmJ (\bmx)\, , \hspace*{2.0in} (A3) \\
\hspace*{1.0in} B^{ij}_{\rm in} (\bmq) &=& \int d^3 x  \int d^3 y \, e^{i  
\bmq \cdot (\bmx - \bmy)} B^{ij}_{\rm in} (\bmx , \bmy)\, ,
\hspace*{1.0in} (A4)
\end{eqnarray*}
and ``$\bot$'' signifies contraction with respect to $\delta^{ij} -  
\hat{q}^{i} \hat{q}^{j}$ (i.e., $\bmJ^2_{\bot} \equiv J^i J^j  
(\delta^{ij} - \hat{q}^i \hat{q}^j)$).  Gauge invariance of this nuclear  
Compton amplitude (discussed in Appendix B) restricts $B^{ij}$ to the 
``inelastic'' part, $B^{ij}_{\rm in}$ (see Eq.\ (B12)).

Because nuclear momentum, size, and energy scales are given by $Q  
\sim 1/R\sim$ 100 MeV, $\omega_N \sim Q^2/2M \sim$ 5 MeV,  
while $m_e \sim$ 0.5 MeV, there are two small dimensionless  
expansion parameters:  $\delta \sim \omega_N R \sim 1/20$ and  
$\beta \sim m_e R \sim  1/200$.  Consequently, we work in configuration  
space, where expansions in R are easiest, using techniques introduced in  
electromagnetically-induced heavy-ion reactions\cite{17}.  Similar techniques  
were used in Refs.\ \cite{6,8}. We introduce the inelastic transition densities  
(squared)
\begin{eqnarray*}
\hspace*{1.5in}\rho_{\rm in} (\bmx , \bmy) &=& \langle 0|  
\rho^{\dagger} (\bmx) | N \rangle \langle N | \rho (\bmy) | 0 \rangle \, ,  
\hspace*{1.5in} (A5)\\  \\
\hspace*{1.25in}J^{ij}_{\rm in} (\bmx , \bmy) &=& \langle 0 | J^{i \dagger}  
(\bmx) | N \rangle \langle N | J^j (\bmy) | 0 \rangle\, ,  
\hspace*{1.37in} (A6)
\end{eqnarray*}
and $B^{ij}_{\rm in} (\bmx , \bmy)$ has already been used in Eq.\ (A4).   
We are not interested in hyperfine structure and, consequently, the spin  
average over nuclear ground-state ($| 0 \rangle$) azimuthal quantum  
numbers is assumed, while the sum over these quantum numbers for  
the intermediate states $(| N \rangle)$ is implicit in Eq.\ (A1).   
Consequently, $\rho_{\rm in}$ and $J^{ii}_{\rm in}$ are space scalars.  The  
plane waves in Eqs.\ (A2 - A4) can then be extracted, appropriate  
spherical averages over $\hat{\bmq}$ performed (e.g., $e^{i \bmq \cdot  
(\bmx - \bmy)} \rightarrow \sin (q z)/(qz)$, where $\bmz \equiv \bmx  
- \bmy$), and the $q$-integrals in Eq.\ (A1) finally evaluated.  This  
produces the generic result
\begin{eqnarray*}
\hspace*{1.00in} \Delta E_{pol} &=& -8 \alpha^2 m_e | \phi_n (0) |^2  
\int d^3 x \int d^3 y \left[ \sum_{N \neq 0} \left[ \rho_{\rm in} 
(\bmx , \bmy) I_N (z) \rule{0in}{2.5ex} \right. \right. \\
&+& \left. J^{ii}_{\rm in} (\bmx , \bmy) J_N (z) + J^{ij}_{\rm in} (\bmx ,
\bmy) z^i z^j \bar{J}_N (z) \right]  \\
&+& \left. \frac{1}{2} B^{ii}_{\rm in} (\bmx , \bmy) K (z) + \frac{1}{2}  
B^{ij}_{\rm in} (\bmx , \bmy) z^i z^j \bar{K} (z) \rule{0in}{4ex} \right] 
\, , \hspace*{0.95in} (A7) 
\end{eqnarray*}
where all of our effort will be devoted to obtaining the polarization  
structure functions:  $I_N (z), J_N (z), \bar{J}_N (z), K (z)$, and  
$\bar{K}(z)$.  We will develop general forms and then perform  
expansions in $(\omega_N z)$ and $(m_e z)$.

We begin with the charge-charge interaction term, $I_N$, which is  
the most complicated to obtain.  All other integrals can be obtained from  
this one:
\begin{eqnarray*}
\hspace*{1.0in} I_N (z) &=& \frac{1}{\omega_N z}  
\int^{\infty}_{0}\frac{dq \, [(2 E + \omega_{N}) \omega_N]\, \sin
(qz)}{q^3 E[(E + \omega_N)^2 - m_e^2]} \\
&=& \frac{1}{\omega_N z} \int^{\infty}_{0} \frac{dq\, \sin (qz)}{q^3  
E} \left[ 1 - \frac{q^2}{(\omega_N + E)^2 - m_e^2} \right] \\
&\equiv& \frac{(I^{(0)} (z) - I^{(1)} (z) )}{\omega_N z}\, .
\hspace*{2.55in} (A8)
\end{eqnarray*}
Nominally infrared divergent, that part of $I^0 (z)$ does not contribute  
after the integrals over $\bmx$ and $\bmy$ are performed in Eq.\ (A7)
(the nucleus  
is assumed to be virtually excited).  Ignoring all of these (ultimately  
vanishing) terms here and in subsequent integrals, we find with $\beta  
\equiv m_e z$
\begin{eqnarray*}
\hspace*{1.10in} I^{(0)} (z) &=& \int^{\infty}_0 \frac{d q \sin  
(qz)}{q^3 E} \\
&=& - \frac{1}{m_e^3} \int^{\beta}_{0} dx \int^{x}_{0} dy  
\int^{y}_{0} d \beta^{\prime}\; K_0 (\beta^{\prime}) \\
&=& - \frac{1}{2 m_e^3} \int^{\beta}_{0} d \beta^{\prime} (\beta -  
\beta^{\prime})^2 \; K_0 (\beta^{\prime})\, , \hspace*{1.45in} (A9)
\end{eqnarray*}
where $K_0 (x)$ is the modified Bessel function of order zero.  The  
function $I^{(0)}(z)$ can easily be expressed in terms of the  
$Ki_n (z)$, the (n\underline{th}) repeated integrals of $K_0 (x)$  
[see Eqs.\ (9.6.25) and (11.2.8) of Ref.\ \cite{38}].  The remaining  
integral $I^{(1)}$ is more challenging and requires several steps.  We  
define $\xi = \omega_N + m_e$ and $\xi^{\prime} = \omega_N -  
m_e$, assuming $\xi^{\prime} > m_e$ for the results reported  
below (this is easily relaxed), and write
\begin{eqnarray*}
\hspace*{1.20in} I^{(1)} &=& \frac{1}{2 m_e} \int^{\infty}_{0}  
\frac{dq}{q} \sin (qz) \left[ \frac{-1}{E(E + \xi)} + \frac{1}{E(E +  
\xi^{\prime})} \right] \\
&\equiv& \frac{ \tilde{I} \, ^{(1)} (\xi^{\prime} ; z) - \tilde{I} \,  
^{(1)} (\xi ; z)}{2m_{e}}\, . \hspace*{2.1in} (A10)
\end{eqnarray*}
Using the field-theory trick\cite{39}
$$
\frac{1}{E_1 E_2 (E_1 + E_2)} = \frac{2}{\pi} \int^{\infty}_0 
\frac{d \lambda}{(\lambda^2 + E^2_1)(\lambda^{2} + E^2_2)}\, , \eqno (A11)
$$ 
and Eq.\ (3.737.3) of Ref.\ \cite{40} the Fourier transforms can be evaluated in  
the form
\begin{eqnarray*}
\hspace*{1.00in} \tilde{I} \, ^{(1)} (\xi ; z) &=& \xi \int^{\infty}_{0}  
\frac{d \lambda}{(\lambda^{2} + \xi^{2})} \frac{(1 - e^{-z  
\sqrt{\lambda^{2} + m^{2}_{e}}}\, )}{(\lambda^{2} + m^{2}_{e})}  
\\
&=& \frac{\pi}{2m_e (\xi + m_{e})}- \frac{\xi}{m_e (\xi^{2} -  
m^{2}_{e})} \left[ \int^{\infty}_{1} \frac{d x\, e^{- \beta x}}{x  
\sqrt{x^{2} - 1}} \right. \\
&-& \left. \int^{\infty}_{1} \frac{d x\, x\, e^{- \beta x}}{(x^{2} +  
\mu^{2}) \sqrt{x^{2} - 1}} \right], \hspace*{1.8in} (A12)
\end{eqnarray*}
after using partial fractions and $z^2 \lambda^2 \equiv \beta^2 (x^2 - 1)$,  
and defining $\mu^2 = \xi^2 / m^2_e -1 (> 0)$.  The first of the remaining  
integrals is $Ki_1 (\beta) = \pi/2 - \int^{\beta}_0 d \beta^{\prime} K_0  
(\beta^{\prime})$ (use $x = \cosh (t)$ in Eq.\ (11.2.10) of Ref.\ \cite{38}).   
The remaining integral (defined to be $\tilde{I} (\beta ; \mu)$ with the
minus sign) yields to a further trick; it is the solution of the  
differential equation
$$
\frac{\partial^2 \tilde{I} (\beta)}{\partial \beta^2} + \mu^2 \tilde{I} 
(\beta) = K^{\prime}_0 (\beta)\, , \eqno (A13)
$$
subject to the boundary condition, $\tilde{I} (\infty) \rightarrow 0$
exponentially, for which standard solutions exist:
\begin{eqnarray*}
\hspace*{.20in} \tilde{I} (\beta ; \mu) &=& - \frac{1}{\mu}  
\int^{\infty}_{\beta} d \beta^{\prime} K^{\prime}_{0}  
(\beta^{\prime}) [\cos (\mu \beta^{\prime}) \sin (\mu \beta) - \cos (\mu  
\beta) \sin (\mu \beta^{\prime})] \\
&=& \frac{-m_e}{\xi} \; [ \sin (\mu \beta) \log (\mu + \sqrt{\mu^{2} +  
1}) + \frac{\pi}{2} \cos (\mu \beta)] \\
&+& \int^{\beta}_{0} d \beta^{\prime} K_0 (\beta^{\prime}) [\sin  
(\mu \beta) \sin (\mu \beta^{\prime}) + \cos (\mu \beta) \cos (\mu  
\beta^{\prime})]\, , \hspace*{.85in} (A14)
\end{eqnarray*}
after integration by parts and rearrangement.  Combining terms from  
Eq.\ (A12) and using the last form of Eq.\ (A9), we obtain  
$\tilde{I}^{(1)} = - \xi I^{(0)} + \bar{I} (\xi ; \beta)$, where
\begin{eqnarray*}
\hspace*{0.10in} \bar{I} (\xi ; \beta) &=& \frac{\xi}{m_e (\xi^{2}  
- m^{2}_{e})} \left\{ -\frac{\pi m_{e}}{2 \xi} ( 1 - \cos (\mu \beta))  
+ \frac{m_e}{\xi} \sin (\mu \beta) \log \left[ \frac{\xi +  
\sqrt{\xi^2 - m_e^2}}{m_e} \right] \right. \\
&-& \int^{\beta}_0 d \beta^{\prime} K_0 (\beta^{\prime})
\left[ \sin (\mu \beta) \sin (\mu \beta^{\prime})
+ \cos (\mu \beta) \cos (\mu \beta^{\prime}) \rule{0in}{3.5ex} \right. \\
&-& \left. \left. 1 - \mu^{2} \beta \beta^{\prime}
+ \frac{\mu^{2}}{2} (\beta^2 + {\beta^{\prime}}^2) 
\right] \right\} \, . \hspace*{2.5in} (A15)
\end{eqnarray*}

Finally, a relatively simple result is obtained
\begin{eqnarray*}
\hspace*{0.1in} I_N (z) &=& \frac{1}{2 m_{e} \omega_{N} z}  
\left[ \bar{I} (\xi ; \beta) - \bar{I} (\xi^{\prime} ; \beta) \right] \\
&\cong& \frac{-z^{2}}{6 \omega_{N}} \left( 1 + \log \left(\frac{2  
\omega_{N}}{m_{e}}\right) + \Delta L_1 \right)
+ \frac{\pi z^{3}}{24}\\
&+&  \frac{\omega_{N} z^{4}}{40} \left(  
\frac{}{} \gamma + \log (\omega_N z) 
+ \frac{\Delta L_{3}}{3} \; - \frac{39}{20} \right) + O (z^5)\, ,  
\hspace*{1.35in} (A16) \\
\end{eqnarray*}
where $\gamma$ is Euler's constant and
\begin{eqnarray*}
\hspace*{.10in} \Delta L_n &=& \frac{\omega_N}{2 m_e} \left[  
\left( \frac{\xi^2 - m_e^2}{\omega_N^2}\right)^{n/2}  
\log \left( \frac{\xi + \sqrt{\xi^2 - m_e^2}}{m_e} \right)  
\right. \hspace*{1.7in} (A17)  \\
&-& \left. \left( \frac{\xi^{\prime 2} -  
m_e^2}{\omega_N^2} \right)^{n/2} \log \left(  
\frac{\xi^{\prime}+ \sqrt{\xi^{\prime 2} -  
m_e^2}}{m_e} \right) \right] 
- \left( 1 + n\, \log \left(\frac{2 \omega_N}{m_e} \right) \right) \\
&\cong& \frac{m_e^2}{\omega_N^2} \left[ a_n \log \left(  
\frac{2 \omega_N}{m_e}\right) + b_n \right] 
+ \frac{m_e^4}{\omega_N^4} \left[ \bar{a}_n \log 
\left( \frac{2 \omega_N}{m_e} \right) + \bar{b}_n \right] + O  
\left( m_e^6/ \omega_N^6 \right)\, .   
\end{eqnarray*}
Useful values are $(a_1 = 1/2, b_1 = -5/12)$, $(\bar{a}_1 = 7/8,  
\bar{b}_1 = - \frac{449}{480})$, and $(a_3 = -\frac{1}{2}, b_3 =  
\frac{1}{12})$.  This completes treatment of the charge term.

The current and seagull terms require special treatment because of the  
transverse projection (i.e., the ``$\perp$'').  The appropriate  
current-current terms have the generic form
\begin{eqnarray*}
\hspace*{0.25in} \int d^3 q\, f (q) | \langle N| \bmJ_{\perp} (\bmq) | 0  
\rangle |^2 &=& \int d^3 x \int d^3 y\, J^{ij}_{\rm in} (\bmx, \bmy) \int  
d^3 q\, f (q) \left( \delta^{ij} - \frac{q^{i} q^{j}}{q^{2}} \right) 
e^{i \bmq \cdot \bmz} \\
&=& \int d^3 x \int d^3 y \rule{0in}{4.0ex} \left\{ J^{ii}_{\rm in} 
(\bmx , \bmy) \int d^3 q\, e^{i \bmq \cdot \bmz} f (q) \right. \\
&+& \left. J^{ij}_{\rm in} (\bmx , \bmy )\; \nabla^i_z \nabla^j_z \int 
\frac{d^{3} q\,  
e}{q^{2}}^{i \bmq \cdot \bmz} f (q) \right\}. \hspace*{0.8in} (A18)
\end{eqnarray*}  
The second q-integral (defined to be $J_{CG}$) can be seen from Eq.\ (1) 
to be \\
$\frac{1}{4 m_e^2}\left[ I_N (z) - \lim_{m_e \rightarrow 0} I_N (z) \right]$,
while the first integral (defined to be $J^0$) is given by
[$- \frac{1}{4 m_e^2 z} \frac{d^{2}}{dz^{2}} z \left[ I_{N} (z) 
- \lim_{m_e \rightarrow 0} I_N (z) \right]$]. Both
forms have an infrared divergence (which will ultimately disappear because
of gauge invariance) and we write $m_e \rightarrow  
2 \lambda$ in divergent logarithms, where $\lambda$ is the equivalent  
small-$q$ cutoff in the integrals.  It is easy to perform the  
derivatives, take the limits, and perform the subtractions. For the sake of 
brevity, we quote only the power-series forms.
$$
J^0 \cong \frac{1}{4 m_e^2\, \omega_N} \left[ 
\log (2 \lambda / m_e) + \Delta L_1  -  
\frac{\omega^2_N z^2}{6} \Delta L_3 \right] + O (z^3) \, ,
\eqno (A19)
$$
$$
J_{CG} \cong - \frac{z^2}{24 m^2_e\, \omega_N} \left[ \log (2 \lambda/ m_e )
+ \Delta L_1 - \frac{\omega_N^2 z^2 \Delta L_3}{20} \right] + O (z^5) \, .
\eqno (A20)
$$
Performing the derivatives in Eq.\ (A18), and matching to Eq.\ (A7),  
we find
$$
J_N \cong \frac{1}{6 m_e^2\, \omega_N}
\left[ \log (2 \lambda / m_e)  + \Delta L_1  - \frac{\omega_N^2 z^2}{5} 
\Delta L_3 \right] \eqno (A21)
$$
and
$$
\bar{J}_N \cong \frac{\omega_N \Delta L_3}{60 m_e^2}\, . \eqno (A22)
$$
The seagull integrals are straightforward variants of Eq.\ (A9).  We  
find
$$
K \cong - \frac{\log ( 2 \lambda / m_e)}{6m_e^2} + \frac{z^2}{60} 
(\gamma + \log (\beta / 2) - 77/60) + O (z^4)\, , \eqno (A23) 
$$
and
$$
\bar{K} \cong - \frac{1}{120} (\gamma + \log (\beta / 2) - 31/30) 
+ O (z^2)\, . \eqno (A24)
$$ 

We have kept terms $\sim R^4$ with the charge densities, since the  
term of $O(1)$ leads to a vanishing result and therefore the leading order 
$\sim R^2$.  The currents themselves are of leading order $R^2$, so  
we have kept terms through $R^4$ (i.e., of relative order $R^2$), as 
in the charge-density case.

With hindsight, we can identify the dominant terms in the expansion,  
which are thoroughly discussed in the main text.  We note the  
appearance of $\log (z)$ terms and a $z^3$ term.  If the Fourier  
transforms of $\rho$ and $J$ could be term-by-term expanded in  
$q^2$, these terms would be absent.  The integrals diverge at some  
order, signaling this with ``non-analytic'' terms in $z^2\, (e.g., z^3 =  
(z^2)^{3/2})$.  Such terms play an important role in modern effective field 
theories\cite{41}. In small terms that we will only estimate, $z^4$ charge  
terms and $z^2$ seagull terms, we will replace $\log (z)$ and $\log  
(\omega_N)$ by average (constant) values, $\log (\bar{z})$ and $\log  
(\bar{E})$, respectively, in order to obtain tractable expressions  
for estimation.

A tedious application of Cartesian moments (discussed in Appendix B)  
after expanding $z^2$ and $z^4$ in powers of $\bmx$ and $\bmy$ leads to
$$
\Delta E_n = -8 \alpha^2 m_e | \phi_n (0)|^2 [A + B +  C]\, , \eqno(A25)
$$
where the charge-charge contribution is
\begin{eqnarray*}
A &=& \sum_{N \neq 0} \frac{|\langle N| \bmD |0 \rangle |^2}{3  
\omega_N} [1 + \log (2 \omega_N / m_e) + \frac{m_e^2}{2 \omega_N^2} 
( \log (2 \omega_N / m_e ) - 5/6)] \\
&+& \frac{\pi}{24} \int d^3 x \int d^3 y\, |\bmx - \bmy|^3\, 
\rho_{\rm in} (\bmx , \bmy ) 
+ \sum_{N \neq 0} \frac{\omega_N}{10} (\gamma + \log (\bar{E} \bar{z}) 
- \frac{39}{20}) [| \langle N | Q^{ij} | 0 \rangle |^2 \\
&+& \frac{1}{2} |  
\langle N | r^2 | 0 \rangle |^2 - 2 \langle 0 | \bmD | N \rangle \cdot \langle  
N | \bmO | 0 \rangle ]\, , \hspace*{2.4in} (A26)
\end{eqnarray*}
while the current-current contribution is
\begin{eqnarray*}
B &=& \sum_{N \neq 0} \frac{| \langle N | \bmD | 0 \rangle |^2}{12  
\omega_{N}} \left( \frac{2 \omega_N^2}{m_e^2} \log (2 \lambda / m_e) +
 \log (2 \omega_N / m_e) - 5/6  \right. \\
&+& \left. \frac{7 m_e^2}{4 \omega_N^2} \left( \frac{}{}  
\log (2 \omega_{N} / m_e) - \frac{449}{420} \right) \right) - \sum_{N  
\neq 0} \frac{\left( \log (2 \omega_{N} / m_{e}) - 1 / 6 \right) | \langle N  
| \bmmu | 0 \rangle |^{2}}{12 \omega_{N}} \\
&+& \sum_{N \neq 0} \frac{(\log (2 \omega_{N} / m_{e}) - 1/6)}{240}  
\left( \frac{4}{3} \omega_N \langle 0 | \bmO | N \rangle \cdot \langle  
N | \bmD | 0 \rangle - \frac{3}{2} \omega_N | 
\langle N | \bar{Q}^{ij} | 0 \rangle |^2 \right. \\
&-& \left. \frac{20 i}{3} \left( \langle 0 | \bmN | N \rangle \cdot \langle  
N | \bmD | 0 \rangle - \langle 0 | \bmD | N \rangle \cdot \langle N |  
\bmN | 0 \rangle \right) \right), \hspace*{1.55in} (A27) \\
\end{eqnarray*}
where $\bar{Q}^{ij}$ is the traceless (E2) part of $Q^{ij}$, while the 
seagull term is
\begin{eqnarray*}
C &=& \int d^3 x \int d^3 y \left[ - \frac{B^{ii}_{\rm in} (\bmx , \bmy)}
{12 m_e^2} \log (2 \lambda / m_e) \right. \\
&+&  \frac{(\gamma + \log (\beta/2) - 4/3)}{8} \left[ \frac{- (B^{ii}_{\rm in}
(\bmx , \bmy) \bmx \cdot \bmy - B^{ij}_{\rm in} x^{i} y^{j})}{12}\right] \\
&+& \frac{1}{240} \left[ \left[ \gamma + \log (\beta/2)  \right] 
\left( B^{ii}_{\rm in}(4 y^2 - \frac{3}{2} \bmx \cdot \bmy ) +
B^{ij}_{\rm in} (- 2y^i y^j - \frac{3}{2} x^i y^j + y^i x^j) \right) \right. \\
&-&  \left. \left. \frac{1}{30} \left(B^{ii}_{\rm in}(154 y^2 - 54 \bmx \cdot  
\bmy) + B^{ij}_{\rm in} (- 69 x^i y^j - 62 y^i  y^j + 31 y^i x^j) \right)  
\right] \rule{0in}{3.5ex} \right]\, , \hspace*{0.25in} (A28) \\
\end{eqnarray*}
where we have separated out a special seagull term (the second term in 
brackets) that has the multipole character of M1$^2$ and defines the 
diamagnetic susceptibility\cite{21}:
$$
\chi_D (\bmx , \bmy ) = - \frac{1}{12} (B^{ii}_{\rm in} (\bmx , \bmy )\, \bmx  
\cdot \bmy - B^{ij}_{\rm in}(\bmx , \bmy )\, x^i   y^j )\, . \eqno (A29)
$$
We note the small coefficients $( \sim 1/240)$ of the last of the current  
and seagull terms.  Similar terms that arise from the charge are an  
order of magnitude larger, as we will see through the use of the gauge  
sum rules of Appendix B.  The former terms are estimated in the main body of 
the paper and will prove to be entirely negligible. Consequently, we will 
ignore those current and seagull terms in what follows. We also note that the  
potentially large (infrared) factor of $\log (m_e)$ cancels in these two 
higher-order terms, except for the coefficient of $\chi_D$.

Relation (B14) shows that the infrared-divergent terms $\sim \log (2  
\lambda / m_e)$ cancel identically.  Using the relations (B15 - B19), we  
can reexpress the last term in (A26) in terms of seagull operators.  We  
finally obtain
\begin{eqnarray*}
 \lefteqn{A+B+C = \left[ \sum_{N \neq 0} \frac{|  
\langle N | \bmD | 0 \rangle |^2}{3 \omega_{N}} \left( 1 + \log \left( 2  
\omega_N / m_e \right)
+ \frac{m_e^2}{2 \omega_N^2} \left( \log 
\left( 2 \omega_N / m_e \right) - 5/6 \right) \right) \right. } \\
&+&  \frac{\pi}{24} \int d^3 x \int d^3 y\, | \bmx - \bmy |^3  
\rho_{\rm in} \left( \bmx , \bmy \right) +
\frac{1}{10} \left( \gamma + \log \left( \bar{E} \bar{z} \right) 
- \frac{39}{20}\right)\\
&\times& \left. \int d^3 x \int d^3 y \left[ B^{ii}_{\rm in} 
\left( \bmx \cdot \bmy - y^2 \right) + B^{ij}_{\rm in}\left( - 2y^i y^j 
+ x^i  y^j + y^i  x^j \right) \right] \rule{0in}{4ex} \right] \\
&+& \left[ \sum_{N \neq 0} \frac{| \langle N | \bmD | 0 \rangle  
|^2}{12 \omega_{N}} \left( \log \left( 2 \omega_{N} / m_{e} \right) 
- 5/6 + \frac{7 m_e^2}{4 \omega_N^2}  
\left( \log \left( 2 \omega_N / m_e \right) - \frac{449}{420} \right)  
\right) \right. \\
&-& \left. \sum_{N \neq 0} \frac{| \langle N | \bmmu | 0 \rangle  
|^2}{12 \omega_{N}} \left( \log \left( 2 \omega_N / m_e \right) - 1/6  
\right) \right] \\
&+& \left[  \frac{1}{8} \int d^3 x \int d^3 y \left( \gamma + \log \left(  
\beta / 2 \right) - 4/3 \right) \chi_D \left( \bmx , \bmy \right) \right]  .  
\hspace*{1.75in} (A30)
\end{eqnarray*}
It is shown in the main text that only the $\bmD, \bmmu$ and $z^3$ terms 
contribute at the part per thousand level. We have enclosed separately in 
large brackets the charge-charge, current-current, and seagull terms. The last
of the charge-charge terms can also be rewritten in a form that allows
estimation of interaction-dependent (potential) terms:
$$
\Delta A = \frac{1}{80} \int d^3 x \int d^3 y\, z^4
\left( \gamma + \log \left( \bar{E} z \right) - \frac{39}{20}\right)
\langle 0 |\, [\rho (\bmx ),[H_0 , \rho (\bmy )]\,]\, | 0 \rangle 
\, . \eqno (A31)
$$
\pagebreak

\begin{center}
\Large{{\bf Appendix B}}\\
\end{center}

In this appendix we perform a Cartesian multipole decomposition of the 
currents and of the virtual nuclear Compton amplitude\cite{14,15}. This is  
unconventional, but affords us the easiest mechanism to impose the  
constraints of gauge invariance.  The Compton amplitude so decomposed 
yields gauge sum rules that express the total content of gauge invariance 
in the long-wavelength limit\cite{14,15}.  These constraints will be imposed 
on the results of Appendix A.

The nuclear current is conserved, or
$$
\mbox{\boldmath $\nabla$} \cdot \bmJ (\bmx) = -i \left[ H_0 , \rho (\bmx)  
\right]\, , \eqno (B1) \\
$$
where $H_0$ is the {\it internal} Hamiltonian (no recoil), and  
this leads to Siegert's Theorem in the long-wavelength limit\cite{15}:
$$
\int d^3 x\, \bmJ (\bmx) \equiv - \int d^3 x\, \bmx \,
\mbox{\boldmath $\nabla$} \cdot \bmJ (\bmx) = i 
\left[ H_0, \bmD \right]\, , \eqno (B2) 
$$
where
$$
\bmD = \int d^3 x\, \bmx \, \rho (\bmx)\, , \eqno (B3)
$$ 
thus removing the explicit effect of interaction currents  
(meson-exchange currents) from the electromagnetic-interaction  
operator in nonrelativistic order (those currents are  
{\it implicitly} present in $H_0$ in the last form of Eq.\ (B2)).   
We can extend this result by expanding Eqs.\ (A2) and (A3) in a power  
series in $(q^i x^i)$, arranging the Cartesian indices of the $\bmx$ and  
$\bmJ$ according to representations of the permutation group:   
symmetric, antisymmetric, and mixed symmetry.  Just as Eq.\ (B2)  
allows the zero\underline{th} moment of $\bmJ (\bmx)$ to be equated  
to the (time derivative of the) first moment of the charge density, all  
symmetric moments of $\bmJ$ are determined by that density through current  
conservation.  The other-symmetry moments are model dependent and  
solely dependent on the magnetic-moment density, $\bmmu (\bmx)$.   
One finds\cite{15}
$$
\rho (\bmq) = \int d^3 x \, \rho (\bmx)\,
e^{i \bmq \cdot \bmx} \cong  
Z + i \bmq \cdot \bmD - \frac{1}{2} Q^{ij} q^i q^j + \cdots \, , \eqno (B4)
$$

\begin{eqnarray*}
\hspace{1in} \bmJ (\bmq) 
&\equiv& \int d^3 x \, \bmJ (\bmx)\, e^{i \bmq  \cdot \bmx} \\
&\cong& i \left[ H_0, \bmD - \frac{q^{2} \bmO}{30} -  
\frac{\bmq \bmq \cdot \bmO}{15} \right] \\ 
&+& \frac{q^2 \bmN}{3} - \frac{\bmq \bmq \cdot \bmN}{3}
- i \bmq \times \bmmu - \frac{1}{2} \left[ H_0, Q^{ij} q^j \right] +  
\cdots \, , \hspace*{0.55in} (B5)
\end{eqnarray*}
where
$$
\bmO = \int d^3 x \,\bmx \, x^2 \, \rho (\bmx)\, , \eqno (B6)
$$

$$
Q^{ij} = \int d^3 x \, x^i \, x^j \, \rho (\bmx) \, , \eqno (B7) 
$$

$$
\bmmu = \frac{1}{2} \int d^3 x \, \left[ \bmx \times  
\bmJ (\bmx) \right]\, , \eqno (B8)
$$

$$
\bmN = \frac{1}{2} \int d^3 x \, \left[ \bmx \times (\bmx \times  
\bmJ (\bmx)) \right] \, . \eqno (B9)
$$
The first five terms in Eq.\ (B5) define unretarded-E1 $(\bmD)$ and retarded-E1
$(\bmO$ and $\bmN)$ interactions, while $\bmmu$ is the magnetic-dipole 
operator, and $Q^{ij}$ the electric quadrupole tensor, which generates 
E2 (via the traceless E2 tensor, $\bar{Q}^{ij}$) and E0 (via the trace of 
$Q^{ij}$) operators.  Terms in $\bmJ$ proportional to $\bmq$ will  
vanish in our case because of the use of Coulomb gauge. The large contribution  
of meson currents makes it convenient to use this decomposition and  
eliminate as much model dependence as possible.  To the order that we  
are working this model dependence resides in $\bmmu$ (M1) and  
$\bmN$ (retarded E1), and Eq.\ (B5) completely and uniquely  
summarizes the constraints of gauge invariance to this order.  Moments  
of the charge and current densities can be obtained by taking  
derivatives with respect to $\bmq$.

We can also develop the constraints of gauge invariance for the  
Compton amplitude.  This is performed in Ref.\ \cite{14}.  Replacing  
$\bmq$ in Eq.\ (B4) by $\bmq_2$ and $\bmq$ in (B5) by $\bmq_1$,  
the gauge-invariance constraint is
$$
\left[ J^m (\bmq_1) , \rho (\bmq_2) \right] = 
-\frac{q^m_2}{M_t} \rho (\bmq_1) \rho (\bmq_2) +  
q^k_2 B^{km} (\bmq_1 , \bmq_2)\, . \eqno (B10)  
$$
We choose for convenience to divide the nuclear Compton amplitude  
into two separate gauge-invariant parts:  elastic and inelastic.  The  
elastic part is the Compton amplitude for a point particle of mass,  
$M_t$, and charge, $Z$, multiplied by two factors of the nuclear  
ground-state charge form factor.  This requires a seagull operator
$$
B^{ij}_{el} (\bmx , \bmy) = \frac{\delta^{ij}}
{M_{t}} \rho_0 (x) \rho_0 (y) \eqno (B11)
$$
for gauge invariance, while the ``inelastic'' seagull is then given by
$$
B^{ij}_{\rm in} (\bmx , \bmy) = B^{ij} (\bmx ,  
\bmy) - B^{ij}_{el} (\bmx , \bmy)\, , \eqno (B12)\\
$$
where $\rho_0$ is the nuclear ground-state charge density, normalized 
to $ \int d^3 x\, \rho_0 (x) = Z$.
The inelastic amplitude will be gauge invariant if the full amplitude is. 
With this definition, the gauge invariance constraint (B10) becomes
$$
\left[ J^m (\bmq_1), \rho (\bmq_2) \right] = q^k_2 B^{km}_{\rm in}  
(\bmq_1 , \bmq_2) - \frac{q^{k}_{2}}{M_{t}} \left( \rho (\bmq_1) \rho  
(\bmq_2) - \rho_0 (q_1) \rho_0 (q_2) \right) \, , \eqno (B13)
$$
where a spin-averaged ground-state expectation value is implied.
Expanding this in powers of $\bmq_1$ and $\bmq_2$ leads to gauge sum rules.

The simplest sum rule results from a single derivative with respect to  
$\bmq_2$:
$$
\left[ \left[ H_0, \bmD \right], \cdot \bmD \right] =  
- \int d^3 x \int d^3 y \, B^{ii}_{\rm in} (\bmx , \bmy)\, . \eqno (B14) 
$$
One $\bmq_2$ and two $\bmq_1$ derivatives produce
$$
\left[ \left[ H_0, \bmO \right], \cdot \bmD \right]  
= - \int d^3 x \int d^3 y \left[ B^{ii}_{\rm in} y^2  
+ 2 B^{ij}_{\rm in} y^i y^j \right] \, , \eqno (B15)
$$
$$
-i \left[ \bmN , \cdot \, \bmD \right] 
= \frac{1}{2} \int d^3 x \int d^3 y \left[ B^{ii}_{\rm in}\, y^2 
- B^{ij}_{\rm in}\, y^i y^j \right]\, , \eqno (B16)
$$
while one $\bmq_1$ and two $\bmq_2$ derivatives generate
$$
\left[ \left[ H_0,  Q^{ij} \right] , Q^{ij} \right] =
-2 \int d^3 x \int d^3 y \left[ B^{ii}_{\rm in}\, \bmx \cdot \bmy 
+ B^{ij}_{\rm in}\, x^i  y^j  \right] + 8 \frac{\bmD^2}{M_t}\, , \eqno (B17) 
$$
$$
\left[ \left[ H_0, \bar{Q}^{ij} \right] , \bar{Q}^{ij} \right] =
-2 \int d^3 x \int d^3 y \left[ B^{ii}_{\rm in}\, \bmx \cdot \bmy 
+ B^{ij}_{\rm in}\, ( x^i y^j  -\frac{2}{3} y^i x^j ) \right] + 
\frac{20}{3}\frac{\bmD^2}{M_t}\, , \eqno (B18) 
$$
$$
\left[ \left[ H_0,  r^2 \right], r^2 \right] 
= -4 \int d^3 x \int d^3 y \;  B^{ij}_{\rm in}\, y^i x^j + 
4 \frac{\bmD^2}{M_{t}}\, . \eqno (B19)
$$
In accordance with our earlier discussion, recoil terms such as the last  
term in Eqs.\ (B17-B19) should be dropped.  Other relations are  
possible, but are not needed.

The gauge sum rules derived above are rigorous in the nonrelativistic limit. 
They incorporate meson-exchange currents in both currents and seagulls. It is 
expected that such currents could alter the impulse-approximation seagull 
by a factor of two, based on numerical studies\cite{22} of Eq.\ (B14) 
reexpressed in its usual (Thomas-Reiche-Kuhn or f-sum rule) form:
$$
2 \sum_{N \neq 0} \omega_{N} | \langle N | \bmD |  
0 \rangle |^2 = \int d^3 x \int d^3 y \; B^{ii}_{\rm in} (\bmx , \bmy)\, .  
\eqno (B20)
$$

\pagebreak


\begin{thebibliography}{99}

\bibitem{1} K.\ Pachucki, D.\ Leibfried, M.\ Weitz, A.\ Huber, W.\  
Konig, and T.\ W.\ H\"{a}nsch, J.\ Phys.\ B{\bf 29}, 177 (1996) contains
an excellent summary of recent experimental and theoretical progress.

\bibitem{2} B.\ de Beauvoir, {\it et al.\ }, Phys.\ Rev.\ Lett.\ {\bf 78}, 440 
(1997).

\bibitem{3} T.\ W.\ H\"{a}nsch, Invited talk at 12\underline{th}
Interdisciplinary Laser Science Conference, Rochester, N.\ Y., Oct.\ 20,
1996; Efforts are under way to reduce this uncertainty by an order of
magnitude, T.\ W.\ H\"{a}nsch (private communication).

\bibitem{4} G.\ G.\ Simon, C.\ Schmidt, F.\ Borkowski, V.\ H. Walter, 
Nucl.\ Phys.\ {\bf A 333}, 381 (1980).

\bibitem{5} I.\ Sick and D.\ Trautmann, Phys.\ Lett.\ {\bf B 375}, 16 (1996).

\bibitem{6} K.\ Pachucki, D.\ Leibfried and T.\ W. H\"{a}nsch, Phys.\  
Rev.\ A{\bf 48}, R1 (1993); K.\ Pachucki, M.\ Weitz, and T.\ W.\  
H\"{a}nsch, Phys.\ Rev.\ A{\bf 49}, 2255 (1994).

\bibitem{7} W.\ Leidemann and R.\ Rosenfelder, Phys.\ Rev.\ C {\bf  
51}, 427 (1995); Y.\ Lu and R.\ Rosenfelder, Phys.\ Lett.\ B{\bf319}, 7  
(1993); (E) {\bf 333}, 564 (1994).

\bibitem{8} J.\ Martorell, D.\ W.\ L.\ Sprung, and D.\ C.\ Zheng, Phys.\  
Rev.\ C{\bf 51}, 1127 (1995). 

\bibitem{8x} A.\ I.\ Milshtein, I.\ B.\ Khriplovich, and S.\ S.\ Petrosyan,
Zh.\ Eksp.\ Teor.\ Fiz.\ {\bf 109}, 1146 (1996) [Sov.\ Phys.\ JETP {\bf 82},
616 (1996)]. This work was performed in zero-range approximation, which is a
very good approximation for the electric polarizability, and somewhat less
good for the magnetic susceptibility.

\bibitem{9} J.\ Bernab\'{e}u and T.\ E.\ O.\ Ericson, Z.\ Phys.\ A{\bf  
309}, 213 (1983).

\bibitem{10} J.\ L.\ Friar and G.\ L.\ Payne, Phys.\ Rev.\ C {\bf  
in press} (1997).

\bibitem{11}  C.\ Zemach, Phys.\ Rev. {\bf 104}, 1771 (1956).

\bibitem{12} J.\ L.\ Friar, Ann.\ Phys. (N.Y.) {\bf 122}, 151 (1979).
See Appendix D for a discussion of Zemach moments.

\bibitem{13} J.\ L.\ Friar, Czech. J.\ Phys.\ {\bf 43}, 259 (1993); H.\ 
Arenh\"ovel, Czech. J.\ Phys.\ {\bf 43}, 207 (1993).

\bibitem{14} J.\ L.\ Friar, Ann.\ Phys. (N.Y.) {\bf 95}, 170 (1975).

\bibitem{15} J.\ L.\ Friar and S.\ Fallieros, Phys.\ Lett.\ {\bf 114B}, 403 
(1982); J.\ L.\ Friar and S.\ Fallieros, Phys.\ Rev.\ C{\bf 29}, 232  
(1984).

\bibitem{16} J.\ L.\ Friar, Phys.\ Rev.\ C{\bf 16}, 1540 (1977).

\bibitem{17} C.\ J.\ Benesh and J.\ L.\ Friar, Phys.\ Rev.\ C {\bf 48}, 1285 (1993).

\bibitem{18} E.\ L.\ Tomusiak, M.\ Kimura, J.\ L.\ Friar, B.\ F.\ Gibson, 
G.\ L.\ Payne, and J.\ Dubach, Phys.\ Rev.\ C {\bf 32}, 2075 (1985).

\bibitem{19} F.\ W.\ K.\ Firk, in {\it Neutron Capture Gamma-Ray Spectroscopy},
ed. by R.\ E.\ Chrien and W.\ R.\ Kane, (Plenum, New York, 1979), p.245.

\bibitem{20} J.\ L.\ Friar, S.\ Fallieros, E.\ L.\ Tomusiak, D.\ Skopik, 
and E.\ G.\ Fuller, Phys.\ Rev.\ C{\bf 27}, 1364 (1983).

\bibitem{21} J.\ L.\ Friar, Phys.\ Rev.\ Lett.\ {\bf 36}, 510 (1976).

\bibitem{22} H.\ Arenh\"ovel, in {\it Proceedings of the Third International 
School of Intermediate Energy Nuclear Physics}, Verona, Italy, 1981, ed. by
R.\ Bergere, S.\ Costa, and C.\ Schaerf, (World Scientific, Singapore, 1982);
H.\ Arenh\"ovel, Z.\ Phys.\ A {\bf 302}, 25 (1981).

\bibitem{23} B.\ Podolsky, Proc.\ Nat.\ Acad.\ Sci.\ U.S.A.\ {\bf 14},  
253 (1928).

\bibitem{24} S.\ Rosendorff and A.\ Birman, Phys.\ Rev.\ A{\bf 31},  
612 (1985).

\bibitem{25} J.\ L.\ Friar, G.\ L.\ Payne, V.\ G.\ J.\ Stoks, and 
J.\ J.\ de Swart, Phys.\ Lett.\ B{\bf 311}, 4 (1993).

\bibitem{26} V.\ G.\ J.\ Stoks, R.\ A.\ M.\ Klomp, C.\ P.\ F.\ 
Terheggen, and  J.\ J.\ de Swart, Phys.\ Rev.\ C{\bf 49}, 2950 (1994).

\bibitem{27} R.\ B.\ Wiringa, V.\ G.\ J.\ Stoks, and R.\ Schiavilla, Phys.\  
Rev.\ C{\bf 51}, 38 (1995).

\bibitem{28} R.\ V.\ Reid, Ann.\ Phys.\ (N.\ Y.\ ) {\bf 50}, 411 (1968).

\bibitem{29} R.\ Machleidt, K.\ Holinde, and C.\ Elster, Phys.\ Rep.\ 
{\bf 149}, 1 (1987).

\bibitem{30} M.\ LaCombe, {\it et al.\ }, Phys.\ Rev.\ C {\bf 21}, 861
(1980).

\bibitem{31} R.\ de Tourreil, B.\ Rouben,  and D.\ W.\ L.\ Sprung, 
Nucl.\ Phys.\ A {\bf 242}, 445 (1975).

\bibitem{32} R.\ B.\ Wiringa, R.\ A.\ Smith, and T.\ A.\ Ainsworth,  Phys.\
Rev.\ C {\bf 29}, 1207 (1984).

\bibitem{33} M.\ M.\ Nagels, T.\ A.\ Rijken, and J.\ J.\ de Swart, 
Phys.\ Rev.\ D {\bf 17}, 768 (1978).

\bibitem{34} R.\ de Tourreil and D.\ W.\ L.\ Sprung,  Nucl.\ Phys.\ A {\bf
201}, 193 (1973).

\bibitem{35} S.\ Kopecky, P.\ Riehs, J.\ A.\ Harvey, and N.\ W.\ Hill,
Phys.\ Rev.\ Lett.\ {\bf 74}, 2427 (1995).

\bibitem{36} J.\ J.\ de Swart, C.\ P.\ F.\ Terheggen, V.\ G.\ J.\ Stoks,
Nijmegen preprint THEF-NYM-95.11, nucl-th/9509032, Proc.\ of Third 
Int.\ Symposium "Dubna Deuteron 95", Dubna, Russia, July '95;
J.\ J.\ de Swart, R.\ A.\ M.\ Klomp, M.\ C.\ M.\ Rentmeester, Th.\ A.\ Rijken,
Few-Body Systems Suppl. {\bf 99},  (1995) and THEF-NYM-95.08.

\bibitem{37} J.\ D.\ Bjorken and S.\ D.\ Drell,
{\it Relativistic Quantum Mechanics}, (McGraw-Hill, New York, 1964). We use
the metric and conventions of this reference.

\bibitem{38} M.\ Abramowitz and I.\ A.\ Stegun, {\it Handbook of
Mathematical Functions}, (Dover, New York, 1965).

\bibitem{39} J.\ L.\ Friar and S.\ A.\ Coon, Phys.\ Rev.\ C {\bf 49}, 
 1272 (1994); Th.\ A.\ Rijken, Ann.\ Phys.\ (N.\ Y.) {\bf 208}, 253 (1991).

\bibitem{40} I.\ S.\ Gradshteyn and I.\ M.\ Ryzhik, {\it Table of Integrals, 
Series, and Products}, ed.\ by A.\ Jeffrey, (Academic Press, Boston, 1994).

\bibitem{41} J.\ L.\ Friar, Few-Body Systems (to appear).

\end{thebibliography}
\end{document}